# Native Point Defects in $Ti_3GeC_2$ and $Ti_2GeC$

M. H. N. Assadi[1,2,*] and H. Katayama-Yoshida[2]

(Dated 2017)

[1]*International Center for Materials Nanoarchitectonics, National Institute for Materials Science (NIMS), 1-1 Namiko, Tsukuba, Ibaraki 305-0044, Japan*

[2]*Department of Materials Physics, Graduate School of Engineering Science, Osaka University, Osaka 560-8531, Japan*

[*]Email: h.assadi.2008@ieee.org      Tel: +81298513354

Using density functional theory, we calculated the formation energy of native point defects (vacancies, interstitials and antisites) in MAX phase $Ti_2GeC$ and $Ti_3GeC_2$ compounds. Ge vacancy with formation energy of 2.87 eV was the most stable defect in $Ti_2GeC$ while C vacancy with formation energy of 2.47 eV was the most stable defect in $Ti_3GeC_2$. Ge vacancies, in particular, were found to be strong phonon scattering centres that reduce the lattice contribution to thermal conductivity in $Ti_2GeC$. In both compounds, the reported high thermal and electrical conductivity is attributed to the electronic contribution that originates from the high density of states at the Fermi level.

## I. INTRODUCTION

MAX phase materials in which M is a transition metal atom, A is a group A element and X is either carbon or nitrogen have a unique and unusual set of properties [1, 2]. MAX phases exhibit strong metallic properties such as high thermal and electrical conductivities that sometimes exceed the values of their corresponding transition metal elements [3, 4]. MAX phase materials are also machineable and have high thermal shock resistance. At the same time, these materials exhibit ceramic properties too; they are resistant to chemical attack, wear and creep, and they are highly stiff. Consequently, MAX phase materials are excellent candidates for applications where metallic and ceramic properties are simultaneously desired such as in high temperature and high impact environments. One application example is to use these materials as components of internal combustion engines [5].

All MAX phases have a space group P63/mmc (no. 194) which consists of alternating near-close-packed layers of $M_6X$ octahedra sandwiched by layers of group A atoms as demonstrated in Fig. 1. Their mechanical properties can be traced back to the fact that the basal dislocations (dislocations along [001]) easily multiply and are mobile at room temperature [6]. The high electrical conductivity of MAX phases is a consequence of the high density of states (DOS) at the Fermi level in these compounds [7, 8]. One particularly interesting MAX phase family is that of (Ti, Ge, C) compounds. These compounds exhibit high bulk modulus compared to other MAX phases [9, 10]. Both the mechanical and the electrical properties of these compounds are expected to be critically influenced by the presence of the native point defects. Therefore, in this work, we will present a comprehensive theoretical study of the formation of native point defects in the $Ti_2GeC$ and $Ti_3GeC_2$ compounds

## II. SYSTEM SETTINGS

Density functional calculations were carried out using SIESTA code [11, 12] that utilises numeric atomic orbital basis and norm-conserving pseudopotentials [13]. The pseudopotentials contained $2s^2 2p^2$ electrons for C, $3d^2 4s^2 4p^0$ electrons for Ti and $3d^{10} 4s^2 4p^2$ for Ge. For atomic orbitals type and exchange-correlation functional, we chose split-valence double-$\zeta$ basis set [14] and generalised gradient approximation based on Perdew-Burke-Ernzerhof parameterisation respectively [15, 16]. The energy cutoff was set to 450 Ry. Integration over the Brillouin zone was carried out using a $k$-point mesh generated based on Monkhorst-Pack scheme [17]. To ensure high accuracy and consistency, the spacing between the $k$-points was maintained at 0.02 Å$^{-1}$ in all calculations. This separation translated into a $17 \times 17 \times 3$ mesh for the primitive cells (Fig. 1) and a $6 \times 6 \times 3$ mesh for the $3a \times 3a \times 1c$ supercells that were used to calculate the formation energy of defects and vibrational properties. Phonon band structures were calculated using the finite displacement method [18] that required the calculation of the force-constants matrix for which the atomic displacement was set to 0.021 Å (0.04 Bohr). The convergence of total energy with respect to the $k$-point mesh, orbital basis and energy cutoff was examined by slightly varying these values. In all cases, the variation in total energy was less than 10$^{-5}$ eV/atom. Therefore, the computational settings were accurate. To test the convergence of the defect's total energy with respect to the supercell size, we repeated the formation energy calculation of select point defects using a supercell of a $4a \times 4a \times 1c$ dimension. The difference in the formation energy was less than ~ 0.01 eV, indicating good convergence. Vacancy defects were created by removing a single atom from the supercell. For interstitial defects, we considered a few distinct configurations. All interstitial defects were more stable in the Ge plane. Antisite defects were created by swapping two adjacent atoms of different chemical species.

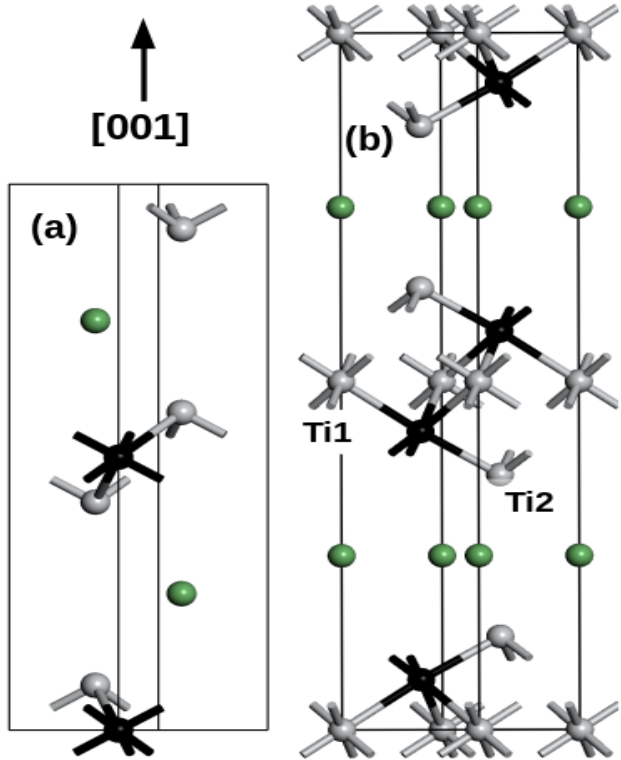

**Fig. 1.** The crystal structure of (a) Ti$_2$GeC and (b) Ti$_3$GeC$_2$ primitive cells. The black, grey and green spheres represent C, Ti and Ge atoms respectively. In Ti$_2$GeC, Ti, Ge and C atoms occupy 4f, 2d, and 2a Wyckoff sites respectively. In Ti$_3$GeC$_2$, Ti1, Ti2, Ge and C atoms occupy 2a, 4f, 2b and 4f Wyckoff sites respectively. In both compounds, the interstitial atoms were stabilised in Ge plane, that is, having a fractional coordinate Z = 3/4 in Ti$_2$GeC and Z =1/4 in Ti$_3$GeC$_2$.

### III. RESULTS AND DISCUSSION

The calculated lattice constants of Ti$_2$CeC were $a$ = 3.10 Å and $c$ = 13.02 Å differing from the experimental values by 1.07% and 0.70% respectively [19]. In the case of Ti$_3$GeC$_2$, the calculated lattice constants were found to be $a$ = 3.11 Å and $c$ = 17.90 Å differing from the experimental values by 1.31% and 0.77% respectively [19]. This slight elongation of lattice parameters is caused by the under-binding character of GGA functional. Nevertheless, the calculated lattice constants were in reasonable agreement with the experimental values. However, in order to eliminate the artificial hydrostatic pressure within the supercell, we conducted all of the electronic and vibrational calculations using the calculated lattice constants. To examine the stability of the MAX phases, we first calculated the 0 K and 0 Pa elemental formation enthalpy ($\Delta E_{elemental}$) of Ti$_2$GiC and Ti$_3$GeC$_2$ with respect to their constituting elements. While 0 K and 0 Pa approximation describes the MAX phases in conditions different from ambient, it has been demonstrated that due to mutual cancellation of the temperature-dependent energy terms of the formation enthalpy in this class of materials, 0 K and 0 Pa approximation can accurately predict MAX phase stability [20, 21]. We found that $\Delta E_{elemental}$ was −4.09 eV/(formula unit) for Ti$_2$GeC and −6.04 eV/(formula unit) for Ti$_3$GeC$_2$. Although, while calculating the formation enthalpies with the desired stoichiometry is trivial, experimentally, due Ge loss in synthesis, the supplied Ge should be % 5 to 10% higher than the nominal value in order to fabricate the desired compounds [22, 23].

We then calculated the stability of Ti$_2$GeC and Ti$_3$GeC$_2$ compounds with respect to their corresponding most competing phases using a screening method based on linear optimisation method [20, 24]. We considered those competing phases that were observed in experimental phase equilibria studies [22] TiC, TiC$_{0.5}$, Ti$_6$G$_5$ and Ti$_5$Ge$_3$. We found that Ti$_2$GeC's most competing phases were Ti$_6$Ge$_5$, and TiC and C related by the following reaction:

$$\text{Ti}_2\text{GeC} \rightarrow \tfrac{1}{5}\text{Ti}_6\text{Ge}_5 + \tfrac{4}{5}\text{TiC} + \tfrac{1}{5}\text{C}. \tag{1}$$

The left-hand side was found to be more stable by 0.362 eV indicating that Ti$_2$GeC is stable with respect to its most competitive phases. In the case of Ti$_3$GeC$_2$, we found that the most competing phases were Ti$_2$GeC and TiC related by the following reaction:

$$\text{Ti}_3\text{GeC}_2 \rightarrow \text{Ti}_2\text{GeC} + \text{TiC}. \tag{2}$$

The left-hand side was more stable by a margin of 0.18 eV indicating that the Ti$_4$GeC$_3$ is also stable with respect to its respective most competitive phases.

The formation energy ($E^f$) of the native point defects was calculated using the following standard formula [25]:

$$E^f = E^t(\text{MAX:D}) + \mu_\alpha - E^t(\text{MAX}) - \mu_M. \tag{3}$$

Here, $E^t(MAX:D)$ is the total energy of the supercell containing defect $D$, and $E^t(MAX)$ is the total energy of the pristine supercell. $\mu_\alpha$ and $\mu_D$ are the chemical potentials of the removed and added elements respectively. The chemical potentials of Ti, Ge, were calculated from the total energies of their respective metallic phases while the chemical potential of C was calculated from the total energy of graphite.

Table 1. The formation energy of the native defects in $Ti_2GeC$. Values in brackets are of those calculations in which only defects' nearest neighbours were allowed to relax.

| Defect | Formation Energy (eV) |
| --- | --- |
| $V_{Ti}$ | 4.50 |
| $V_{Ge}$ | 2.87 |
| $V_C$ | 2.93 |
| $Ti_{Int}$ | 4.40 |
| $C_{Int}$ | 3.05 |
| $Ge_{Int}$ | 2.90 |
| $C_{Ti}$ | 6.26 (6.44) |
| $Ge_{Ti}$ | 4.33 (4.46) |
| $C_{Ge}$ | 6.78 (7.11) |

Table 2. The formation energy of the native defects in $Ti_3GeC_2$.

| Defect | Formation Energy (eV) |
| --- | --- |
| $V_{Ti1}$ | 7.14 |
| $V_{Ti2}$ | 4.85 |
| $V_{Ge}$ | 2.90 |
| $V_C$ | 2.47 |
| $Ti_{Int}$ | 3.73 |
| $C_{Int}$ | 2.67 |
| $Ge_{Int}$ | 3.29 |
| $C_{Ti1}$ | 6.28 |
| $C_{Ti2}$ | 5.81 |
| $Ge_{Ti1}$ | 7.12 |
| $Ge_{Ti2}$ | 4.72 |
| $C_{Ge}$ | 6.97 |

The $E^f$ of defects in $Ti_2GeC$ is presented in Table 1. Among the vacancies, $V_{Ge}$ had the lowest $E^f$ of 2.87 eV while $V_C$ and $V_{Ti}$ had higher $E^f$s of 2.93 eV and 4.50 eV respectively. In the case of interstitial defects, $Ge_{Int}$ had the lowest formation energy of 2.90 eV followed by $C_{Int}$ and $Ti_{Int}$ with $E^f$s of 3.05 eV and 4.40 eV respectively. For antisite defects, $Ge_{Ti}$ had the lowest $E^f$ of 4.33 eV followed by $C_{Ti}$ with an $E^f$ of 6.26 eV and $C_{Ge}$ with an $E^f$ of 6.78 eV. While optimising the structures of the antisite defects, we noticed that atoms surrounding the defects experienced large displacements which resulted in considerable distortions in the lattice structures. For instance, in the case of $C_{Ti}$ some of the atoms in the vicinity of the swapped atoms had a net displacement of ~ 2 Å from their original locations. The magnitude of this distortion is comparable to the Ti–C bond length of 2.14 Å. Such large distortion in the relaxed structure originates from the large difference between the ionic radius of Ti (1.76 Å) and that of C (0.67 Å). In this case, large atomic relaxation can cause an artificial elastic interaction which affects the calculated formation energy of the defects [26]. To examine the extent of this error, we recalculated the $E^f$s of the antisite defects under the constraints that restricted the relaxation to the defects' nearest neighbours while fixing all other atoms to their original sites. This procedure, to a great extent, truncates the artificial elastic interaction [26]. The corresponding $E^f$ values are presented in brackets in Table 1 which are generally ~ 0.3 eV higher than the values obtained for fully relaxed structures. Consequently, the elastic error for antisite defects in $Ti_2GeC$ is indeed an order of magnitude smaller than the formation energies and does not affect the general trend of the defects' stability and the likelihood of their formation. As in $Ti_2GeC$, $V_{Ge}$ was also predicted to be the most stable vacancy in the closely related compound $Nb_2GeC$ [27].

The formation energy of $Ti_3GeC_2$'s defects is presented in Table 2. As demonstrated in Fig. 1(b), there are two distinct Ti sites in $Ti_3GeC_2$'s lattice. Consequentially, there are four types of distinct vacancies and five types of antisite defects in $Ti_3GeC_2$. As demonstrated in Table 2, the most stable vacancy was $V_C$ with an $E^f$ of 2.47 eV followed closely by $V_{Ge}$ with an $E^f$ of 2.90 eV. Both Ti vacancies had considerably higher $E^f$s of 4.85 eV for $V_{Ti2}$ and 7.14 eV for $V_{Ti1}$. Among the interstitials, $C_{Int}$ had the lowest formation energy of 2.67 eV followed by $Ge_{Int}$ with an $E^f$ of 3.29 eV and $Ti_{Int}$ with an $E^f$ of 3.73 eV. The formation energies of antisite defects ranging from 4.72 eV for $Ge_{Ti2}$ to 7.12 eV $C_{Ti2}$ were considerably higher than the formation energies of C and Ge vacancy and interstitial defects. We found that in both compounds the formation energy of both $V_{Ge}$ and $V_C$ was considerably lower than that of $V_{Ti}$. Experimentally this implies that the concentration of $V_C$ and $V_{Ge}$ vacancies would be a few folds higher than those of $V_{Ti}$ [28, 29].

In order to examine the electronic structure, the total and partial density of states of both $Ti_2GeC$ and $Ti_3GeC_2$ compounds were calculated and presented in Fig. 2(a) and (b) respectively. The general aspects of the calculated density of states (DOS) are consistent with previous DFT calculations [9, 30, 31]. Both compounds have a narrow band that is located below the valence band at ~ −18 eV to

~ −16 eV which is mainly occupied by C 2s and Ge 4s electrons. In the valence band that stretches from ~ −12 eV up to the Fermi level in both compounds, Ti 3d states are the dominant states providing delocalised electrons at Fermi level giving rise to strong metallic conduction as experimentally observed in all MAX phase materials. Nonetheless, near the Fermi level, Ti 3d electrons hybridise with Ge 4p and C 2p electrons. Consequently, at the Fermi level, $Ti_2GeC$ has a DOS of 27.05 $eV^{-1}$ composed of 15.49 $eV^{-1}$ d states and 11.56 $eV^{-1}$ p states. Similarly, in $Ti_3GeC_2$, the DOS at the Fermi level is 27.24 $eV^{-1}$ of which 15.49 $eV^{-1}$ is d-like, and 11.75 $eV^{-1}$ is p-like. The coexistence of d and p bands at the Fermi level causes the two-band conduction behaviour that has been widely reported in experiments [8, 32]. In this case, both p and d states provide itinerant carriers for conduction resulting in high carrier density of $1.3 \times 10^{27}$ $cm^{-3}$ for $Ti_2GeC$ and $1.4 \times 10^{27}$ $cm^{-3}$ for $Ti_3GeC_2$ [33]. Since the contribution of p and d states to the Fermi level are comparable, one anticipates that the concentration of electrons and holes to be of the same order. This prediction is in agreement with the measurements carried out on $Ti_2CeC$ thin films [34]. Furthermore, the Seebeck coefficient in these compounds is reportedly extremely low as it is of the order of ~ ±2 μV/K [35, 36]. Although low Seebeck coefficient is a consequence of high carrier concentration, the two-band carrier conduction in these systems further contributes to the low Seebeck coefficient as Seebeck effect of n and p carriers nearly cancel over a wide range of temperatures [37]. Such compensated Seebeck effect has widely been reported for similar MAX phase materials [38, 39].

In order to investigate the effect of the most stable point defects on the vibrational modes, the phonon band structures of pristine $Ti_2GeC$ and defective $Ti_2GeC:V_{Ge}$ were calculated and are presented in Fig. 3(a) and (b) respectively. Likewise, the phonon band structures of $Ti_3GeC_2$ and $Ti_3GeC_2:V_C$ are presented in Fig. 4(a) and (b) respectively. By comparing Fig. 3(a) and Fig. 4(a), we observe that in both compounds, the dispersion pattern of band structure along Γ-M-K-Γ is very similar to the dispersion along A-L-H-A indicating a very small dispersion along $c$ direction (Γ-A). This, in turn, leads to anisotropic thermal conduction with heat conductivity being smaller along $c$ direction [40]. Furthermore, due to the mass difference between C and Ge atoms, both compounds have a wide phonon dispersion gap that separates the optical modes from the acoustic modes. The higher frequency modes are made of vibrational modes of C atoms as they have the lightest mass while the bands below the dispersion gap are made of Ti and Ge vibrational modes. In the case of $Ti_3GeC_2$, Ti1 modes have a higher frequency than those of Ti2 as indicated in Fig. 4(a). In $Ti_2GeC$, the dispersion gap starts at ~ 350 $cm^{-1}$ and terminates at ~ 550 $cm^{-1}$ while in $Ti_3GeC_2$, the band gap starts at ~ 400 $cm^{-1}$ and expands up to ~ 550 $cm^{-1}$.

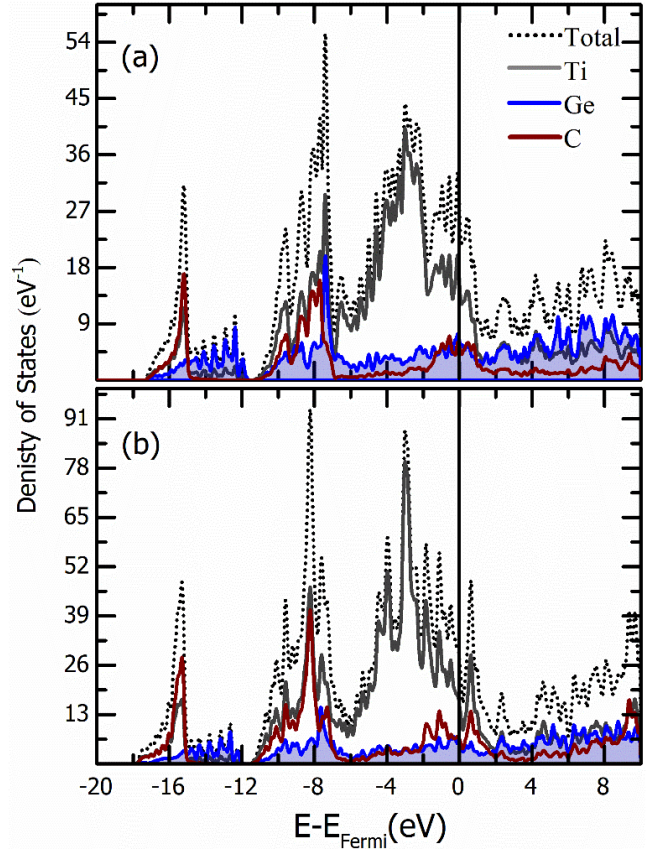

**Fig. 2.** Total and partial density of states of (a) $Ti_2GeC$ and (b) $Ti_3GeC_2$. The dotted, black, red and blue lines represent the total, Ti, C and Ge states respectively.

In the case of defective $Ti_2GeC:V_{Ge}$, as shown in Fig. 3(b), the Ge vacancy the causes a separate set of Ti vibration modes (marked with an arrow in Fig. 3(a)) to disperse into rest of acoustic modes. Separate vibrational modes in the phonon band structure indicate sharp peaks in the phonon density of states which strongly contribute to the thermal conductivity. Therefore, the merging of sharp Ti peaks into the rest of acoustic vibrational modes decreases the thermal conductivity. We, hence, infer that $V_{Ge}$ acts as a potent phonon scatterer. This effect is similar to that of equivalent vacancies in other layered hexagonal structures in decreasing the thermal conductivity [41]. In the case of $Ti_3GeC_2:V_C$, the carbon vacancy causes a flat band at high-frequency region (marked by a green arrow in Fig. 4(b)) indicating that some of the C ions exhibit atom-like behaviour. This peak particularly corresponds to C's vibrations along $c$ direction which indicates that in the $Ti_3GeC_2:V_C$ system, the presence of $V_C$ results in an additional vibrational motion for C atoms. However, since this vibrational mode is of higher frequency, it does not significantly affect the lattice thermal conductivity in $Ti_3GeC_2$.

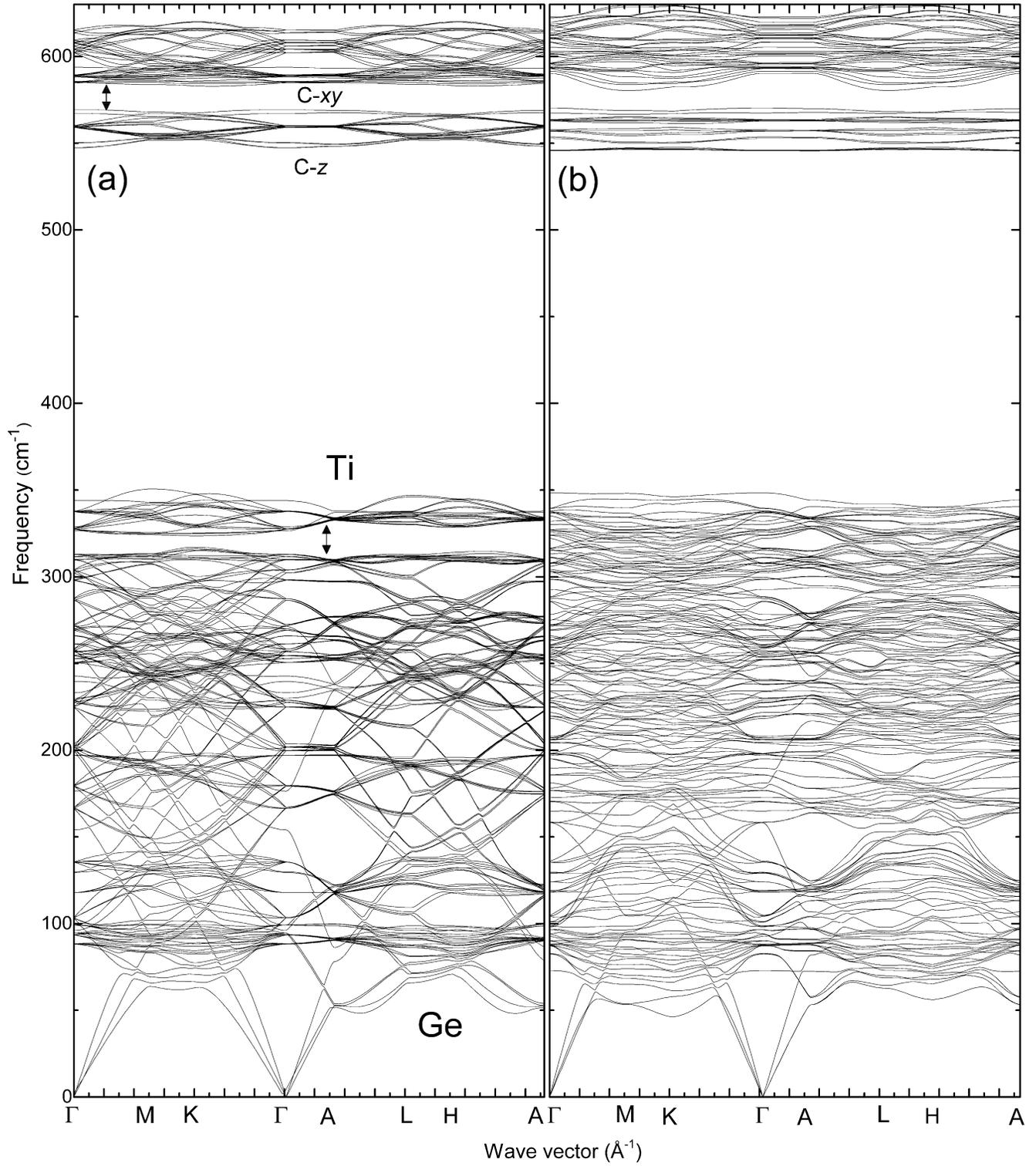

**Fig. 3.** Phonon band dispersion of (a) pristine $Ti_2GeC$ and (b) $Ti_2GeC:V_{Ge}$. Phonon modes were calculated using the supercell approach. No intensity filter was applied. The number of modes is proportional to the number of atoms in the supercell.

M. H. N. Assadi and H. Katayama-Yoshida

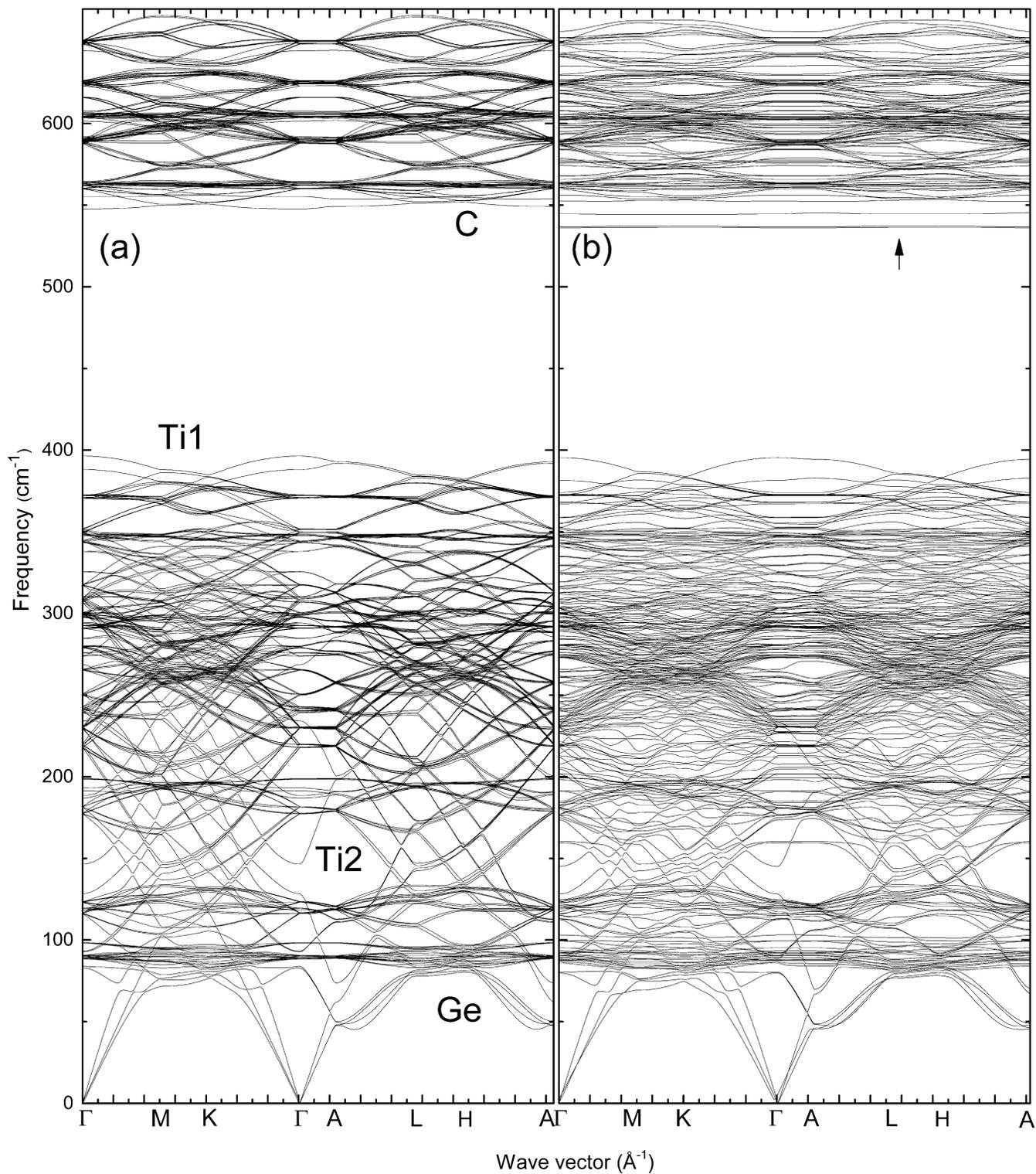

**Fig. 4.** Phonon band dispersion of (a) pristine $Ti_3GeC_2$ and (b) $Ti_3GeC_2$:$V_C$. No intensity filter was applied. The number of modes is proportional to the number of atoms in the supercell.

## IV. CONCLUSIONS

In conclusion, we have found that both Ti$_2$GeC and Ti$_3$GeC$_2$ are energetically stable against the investigated competing phases of TiC, Ti$_{0.5}$C, Ti$_5$Ge$_3$ and Ti$_6$Ge$_5$. Furthermore, V$_{Ge}$ is the most stable native point defect in Ti$_2$GeC while V$_C$ is the most stable native point defects in Ti$_3$GeC$_2$. In both compounds, Ti vacancies and antisite defects have noticeably higher $E^f$s. The p-d band crossing at the Fermi level in both compounds also accounts for the compensated conduction behaviour that has been widely reported by experimentalists. In Ti$_2$GeC, V$_{Ge}$ act as strong phonon scatterer that decreases the lattice thermal conductivity. Therefore, the reported high thermal conductivity in this compound must be attributed to its high electronic contribution.

## ACKNOWLEDGEMENT

This work was supported by The Japanese Society for Promotion of Science.